\documentclass[preprint,11pt]{aastex}

\renewcommand{\deg}{\ifmmode^\circ\else$^\circ$\fi}

\newcommand{\myfigure}[2]{%
\begin{figure}
\plotone{#1}
\caption{#2}
\end{figure}
}

\newcommand{\be}{\begin{equation}}
\newcommand{\ee}{\end{equation}}

\newcommand{\bea}{\begin{eqnarray}}
\newcommand{\eea}{\end{eqnarray}}

\newcommand{\um}{\hbox{$\mu$m}}

\newcommand{\nWmmsr}{\hbox{nW m$^{-2}$ sr$^{-1}$}}
\newcommand{\MJysr}{\hbox{MJy sr$^{-1}$}}
\newcommand{\kJysr}{\hbox{kJy sr$^{-1}$}}

\newcommand{\etal}{{\it et al.\/}}
\newcommand{\vs}{{\it vs.\/}}
\newcommand{\W}{WCVWS}
\newcommand{\E}{EAG}

\newcommand{\Kband}{\mbox{$K$-band}}
\newcommand{\Lband}{\mbox{$L$-band}}

\newcommand{\LkJyval}{14.4}
\newcommand{\LkJy}{\hbox{$\LkJyval \pm 3.7$~\kJysr}}
\newcommand{\KkJy}{\hbox{$16.9 \pm 4.4$~\kJysr}}
\newcommand{\LnW}{\hbox{$12.4 \pm 3.2$~\nWmmsr}}
\newcommand{\KnW}{\hbox{$23.1 \pm 5.9$~\nWmmsr}}
\newcommand{\LGWCkJyval}{12.8}

\begin{document}

\title{Detection of the Cosmic Infrared Background at 2.2 
and 3.5 \um\ Using DIRBE Observations}

\author{E. L. Wright} \affil{Department of Physics and
Astronomy, University of California, Los Angeles, CA  90095-1562}  
\email{wright@astro.ucla.edu}

\and

\author{E. D. Reese} \affil{Department of
Astronomy \& Astrophysics, University of Chicago, 5640 S. Ellis Ave.,
Chicago, IL 60637}
\email{reese@piglet.uchicago.edu}

\begin{abstract}

We compare data from the Diffuse InfraRed Background Experiment (DIRBE) 
on the Cosmic Background Explorer ({\sl COBE}) satellite to the
the \citet{WCVWS92} model of the infrared sky.
The model is first compared with broadband $K$ (2.2 $\mu$m) 
star counts.  Its success at \Kband\ gives credence to its physical approach 
which is extrapolated to the \Lband\ (3.5 $\mu$m).
We have analyzed the histograms of the pixel by pixel intensities
in the 2.2 and 3.5 \um\ maps from DIRBE after subtracting the zodiacal light.
The shape of these histograms agrees quite well with the histogram shape
predicted using the \citet{WCVWS92} model of the infrared sky, but the
predicted histograms must be displaced by a constant intensity in order to
match the data.  This shift is the cosmic infrared background, which is
\KkJy\ or \KnW\ at 2.2 \um, and \LkJy\ or \LnW\ at 3.5 \um.
\end{abstract}


\section{Introduction}

Theoretical consideration of the cosmic infrared background (CIRB) 
radiation began 
decades ago with a paper by \citet{PP67}.  It took quite some 
time for the IR detector technology to catch up to the sensitivity demands in 
detecting the faint CIRB.  The Diffuse InfraRed Background Experiment (DIRBE) 
on the Cosmic Background Explorer ({\sl COBE}) is 
designed specifically to detect the CIRB.  The DIRBE team reported detections
of the CIRB at 140 and 240 \um, and upper limits at other wavelengths from 1.25
to 100 \um, in \citet{HAKDO98}, \citet{KWFRA98} and \citet{AOWSH98}.  
\citet{DA98} reported a {\em lower limit} at 3.5 \um\ based
on the correlation between the 2.2 and 3.5 \um\ DIRBE maps and a lower limit
at 2.2 \um\ based on galaxy counts.  \citet*{GWC00} have reported a tentative
detection of the CIRB at both 2.2 and 3.5 \um\ after directly measuring and
subtracting the galactic stars in a ``dark spot''.

The main source of the CIRB is thought to be the superimposed
light of distant galaxies.  The optical 
and UV radiation emitted from early star formation in such galaxies is 
red shifted to the NEAR INFRARED (NIR) in the present epoch.  
The observation of the CIRB is complicated 
by local contributors to the NIR flux:  scattering and thermal re-emission
from interplanetary dust 
(IPD), known as zodiacal 
light (ZL); the interstellar medium (ISM) in the Milky Way;
and foreground stars in our own galaxy.  
Extragalactic sources include pregalactic 
stars, cosmic explosions, and possibly decaying elementary particles if 
they attain a sufficient density.  
{\it Due to these sources,}
the CIRB offers information regarding the nature and evolution of early 
luminous matter, be it stars or primeval galaxies; limits on the epoch of 
galaxy formation; the presence or absence of dust in early galaxies;
and constraints on the reheating of the universe between $z=5$ and $10^3$ 
\citep{FM93}.  There is a minimum in IPD flux and zodiacal 
light at 3.5 \um\ (see Figure~1 of \citet{Ca92}) creating a `window' for 
observing the CIRB.  The regions away from the galactic plane offer the 
lowest star density, and thus low  contaminating stellar flux, 
plus minimal dust obscuration from our own galaxy.

In this paper we first remove the 
strong zodiacal light foreground and the small interstellar medium foreground
from the DIRBE maps at 2.2 \& 3.5 \um.  The maps after these subtractions
are very strongly confusion limited by the overlapping signals from
galactic stars.  We have modeled the histogram of pixel values using
a modified version of the \citet[hereafter \W]{WCVWS92} IR model of 
the Galaxy to predict star counts at 2.2 \&
3.5~\um, after testing the model against actual observations in several
parts of the Galaxy at 2.2~\um.
The model histograms match the observed histograms very well after a constant
offset is added to the model, and this offset is the value of the CIRB.

\section{Foregrounds}

\subsection{Zodiacal Light}

\citet{KWFRA98} describe the zodiacal light model used by 
\citet{HAKDO98}.  
But \citet{KWFRA98} leave a large residual intensity in the galactic
polar caps at 25 \um, the DIRBE band that is most dominated by the zodiacal
light.  For example, the 25 \um\ intensity toward the DIRBE dark spot at
$(l,b) = (120.8^\circ, 65.9^\circ)$ in the DIRBE ZSMA 
(Zodi Subtracted Mission Average) maps prepared using the
\citet{KWFRA98} model is 1.76 \MJysr.
This cannot be a cosmic background because the lack of $\gamma$-ray emission
toward Mkn 501 limits the CIRB to be $< 33$ \kJysr\
\citep{FMM98}.
It also cannot be a galactic signal because the 100 \um\ intensity in this
field is 1.27 \MJysr\ in the ZSMA maps, and \citet{AOWSH98} specify
the ISM intensity as $R(\lambda)(I(100)-I_\circ)$, with 
the ratio of 25 \um\ to 100 \um\ galactic signals (in \MJysr)
given by $R(25) = 0.0480$, and the 100 \um\ intensity extrapolated to
zero $N_H$ is $I_\circ = 0.66$~\MJysr, 
so the ISM intensity at 25 \um\ is only 29~\kJysr.  By elimination,
most of this intensity must be zodiacal.

In order to reduce the residual zodiacal emission in the maps, \citet{Wr97}
added one ``observation'' that the high $b$ intensity at 25 \um\ should be zero
to the more than $10^5$ observations used in the zodiacal model fitting.  
Even this
very low weight pseudo-observation lowered the 25 \um\ intensity in the dark
spot to 0.26 \MJysr.  This indicates that the isotropic component of the
zodiacal emission is very poorly constrained in fits that just look at the
time variation to measure the zodiacal light.  
The Appendix in \citet{Wr98} discusses zodiacal
light models in more depth, and \citet{GWC00} give the 
actual parameters of the model we have used.

Since the residual 25 \um\ intensity is now only 1\% of the total zodiacal
emission, we might hope for errors in the ZL model equal to 1\% of the ecliptic
pole intensity.  But the situation is more uncertain at 2.2 and 3.5 \um\ due
to the scattered component of the ZL.  Adjusting the thermal emission component
to fit the 25 \um\ intensity will not necessarily lead to a correct scattered
component.  So we have adopted ZL modeling errors of 5\% of the intensity at
the ecliptic poles at 2.2 and 3.5 \um.  These errors are slightly lower at 2.2
\um\ 
(5.2 {\it vs.} 6 \nWmmsr) and higher at 3.5 \um\ 
(2.8 {\it vs.} 2 \nWmmsr) than the errors adopted by \citet{KWFRA98}.

\subsection{Interstellar Medium}

\citet{AOWSH98} give a method for removing interstellar medium emission
from the maps after the zodiacal light is removed.  The corrected map is
given by
\be
I^\prime_\nu(\lambda) = I_\nu(\lambda) - R(\lambda)(I_\nu(100)-I_\circ)
\ee
where the intensity ratio $R(3.5) = 0.00183$ but $R(2.2) = 0$.  $I_\circ$ is
the intercept at zero $N_H$ of the zodi-subtracted 100 \um\ map.  Thus
$I_\circ$ is an intensity that is not from the solar system and not from the
Galaxy.  
In the \citet{AOWSH98} analysis it is assumed to be a cosmic background,
but \citet{HAKDO98} do not claim a detection of a cosmic background at 100 \um\
because of the uncertainties in the analysis that leads to $I_\circ$.
\citet{AOWSH98} give the value $I_\circ = 19.8 \pm 2.5\;\nWmmsr$, or
660 \kJysr.

In our analysis we have used a map of $I_\nu(3.5) - R(3.5) I_\nu(100)$.
Comparing this to the \citet{AOWSH98} formula shows that we need to add
$R(\lambda) I_\circ$ to our final result because we have over-subtracted the
interstellar medium in our maps.  But we should not use the 
$I_\circ = 0.66\;\MJysr$ of \citet{AOWSH98} because we have subtracted more
zodiacal light than the \citet{KWFRA98} model, and this will lower the 
zero $N_H$ intercept in the zodi-subtracted 100 \um\ map.  The difference in
the zodiacal light model averages to $0.3\;\MJysr$ in the regions we study in
this paper, so we use $I_\circ = 0.4\;\MJysr$, and add 0.7 \kJysr\ to our 
3.5 \um\ results.  No correction is needed at 2.2 \um.

\section{The Starcount Model}
\label{sec:model}

Star count calculations used a modified version of the \W\ starcount model
which incorporates the spiral arm modifications made by \citet{Co94}.
Except for recoding and a different treatment of the Gaussian distribution of
absolute magnitudes, our starcount model is the same as that used by
\citet{AOWSH98} for computing their Faint Source Model.
The model breaks the galaxy into five components: disk, spiral arms, molecular
ring, central bulge, and extended halo.  Integration is performed along the
line of sight extending from the Sun to the desired galactic longitude and
latitude through each galactic component.  Each component is populated by 87
different stellar types distributed among the five galactic components using a
weighting technique to simulate the distribution observed in the Galaxy.
The spectral classes are characterized by absolute magnitudes in various IR 
wavelengths, a magnitude dispersion, $z$-component disk scale height 
($z$ is the height above the galactic plane), 
solar neighborhood density, and relative 
weights for each galactic component.
Dust is modeled with a smooth exponential distribution following 
the \citet{RL85} extinction law.  
The model is able to perform calculations in the
$J$, $K$, $L$, and $M$ bands and at 12 and 25 $\mu$m.  
% A summary of the model's density profiles and relevant parameters can be 
% found in Table~\ref{tab:model_sum}, at the end of this paper.

\subsection{Components of the model}

The differential star counts are given by
\bea
\frac{dN}{dm}(m) & = & \int_0^S \left\{
\rule{0in}{4.12ex}
\sum_{i=1}^{87} \rho_\circ^i \left[ 
\exp(-|z|/h_z^i) \left(D(\vec{r})\left\{f_D^i + f_A^i A(\vec{r})\right\} + 
f_R^i {\cal R}(\vec{r}) \right) + f_B^i B(\vec{r}) + f_H^i H(\vec{r})
\right] \right. \nonumber \\
& \times &
\left. \frac{\exp[-(M_\lambda^i+5\log(s/\mbox{10 pc}) + 
A_\lambda(\vec{r}) - m)^2/(2\sigma_i^2)]}{\sigma_i \sqrt{2\pi}}
\rule{0in}{4.12ex}
\right\} s^2 ds
\eea
where the spatial components for the disk, $D(\vec{r})$;
spiral arms, $A(\vec{r})$; molecular ring, ${\cal R}(\vec{r})$;
bulge, $B(\vec{r})$; halo, $H(\vec{r})$; and dust,
$A_\lambda(\vec{r})$ are discussed in the following paragraphs.
The weights $f_D^i$, $f_A^i$, $f_R^i$, $f_B^i$ and $f_H^i$ give the relative 
importance of the disk, arm, ring, bulge and halo spatial components
for the $i^{th}$ star type.
The position along the line-of-sight is given by
\be
\vec{r} = (x,y,z) = (-R_\circ + \cos(l)\cos(b) s, \sin(l)\cos(b) s, 
\mbox{18 pc} + \sin(b) s).
\ee
We use an offset of the Sun above the plane of 18 pc \citep{AOWSH98},
while \citet{Co95} gives 15.5 pc and \citet{HL95} give 20.5 pc.
Note that the spatial components and the volume element $s^2 ds$
do not depend on the star type $i$ and can be precomputed for a given 
line of sight before doing the loop over star types.  This greatly speeds 
up the calculation of the model.  
The upper limit $S$ is set at a radius of 15~kpc from the Galactic center.
The local density $\rho_\circ$,
the vertical scale height $h_z$, the absolute magnitudes $M_\lambda$ and their
standard deviation $\sigma_i$, and the weights $f_D, f_A, f_R, f_B$ and $f_H$
all depend on the star type indexed by $i$.  These parameters are taken from
Table 2 in \W, except for the absolute magnitudes at 3.5 \& 4.9 \um, 
for which we use the extrapolated values from \citet{AOWSH98}.

By directly computing the differential counts, we have no need of the 5 point
approximation to the Gaussian used by \W\ for the absolute magnitude
distribution.  The exact Gaussian gives larger source counts than the 5 point
approximation by an amount ${\cal O}(10^{-1}\sigma^2)$ which amounts to about
10\% for the lower main sequence stars that dominate the faint high latitude
counts.

The disk is assumed to be an exponential disk,
\be
D(\vec{r}) = \exp\left[(R_\circ-R)/h_R\right]
\ee
where $R = \sqrt{x^2+y^2}$ is the radius in cylindrical coordinates.
The $z$ dependence has been factored out since it varies with star type.
The radial scale is $h_R = 3.5$~kpc and the solar position is at
$R_\circ = 8.5$~kpc.

The spiral arms have the same exponential radial dependence as the disk, so
this has been factored out.  The spiral arm function has only two values
corresponding to being in or out of an arm.  Define
$\theta$ such that $x = -R\,\cos\theta$ and
$y = -R\,\sin\theta$.
Thus $\theta = 0$ at the solar position.
The arms are defined as logarithmic spirals, based on the H~II observations
of \citet{GG76}, starting at $R_{min}$ and
$\theta_{min}$ and extending over an angular extent of $E_i$ but
having a fixed radial width of $W_i$.
Then the arm function is given by
\be
A(\vec{r}) = C_{arms} \prod_{j=1}^6
T(|R_{min_i} \exp[(\theta - \theta_{min_i})/\alpha_i] -
 R|
< W_i/2)\;T(\theta - \theta_{min_i} < E_i)
\ee
where $\theta-\theta_{min}$ is taken modulo $2\pi$
and the truth function $T()$ is 1 if its
argument is true, and 0 for false arguments.
The arm normalization is given by $C_{arms} = 5$.
The parameters for the six arms are given in Table~\ref{tab:spiral}.
Figure~\ref{fig:spiral} shows the pattern of the spiral arms and the molecular
ring in the model.
Note that the Sun is just barely within the sixth arm.

The halo function, taken from the \citet{Yo76} approximation to the 
deprojected de~Vaucouleurs $r^{1/4}$ intensity law, is given by
\be
H(\vec{r}) = C_{halo} \frac{\exp(-b\xi)}{\xi^{7/2}\sqrt{32b}}
\ee
with $b = 7.66924944$, and
\be
\xi = \left[\left(\frac{x}{x_H}\right)^2+\left(\frac{y}{y_H}\right)^2+
\left(\frac{z}{z_H}\right)^2\right]^{1/8}
\ee
with $x_H = y_H = 2.83$~kpc, and $z_H = 2.264$~kpc.  The normalization is set
so $H = 0.002$ at the solar position.

The molecular ring density enhancement is radially Gaussian with the same
stellar type dependent vertical scale height as the disk, so
\be
R(\vec{r}) = C_{ring} \exp(-0.5(R-R_{ring})^2/\sigma_{ring}^2)
\ee
The radial position of the ring is at $R_{ring} = 0.45 R_\circ$ and the
radial width is given by $\sigma_{ring} = 0.064 R_\circ$, and the ring
normalization is $C_{ring} = 25$.

The bulge function, proposed by \citet{Ba86} and chosen by \citet{Co94},
is given by
\be
B(\vec{r}) = C_{bulge} \frac{\exp(-\chi^3)}{\chi^{1.8}}
\ee
with
\be
\chi = \left[\left(\frac{x}{x_B}\right)^2+\left(\frac{y}{y_B}\right)^2
+\left(\frac{z}{z_B}\right)^2\right]^{1/2}
\ee
with bulge scales $x_B = y_B = 2$~kpc, $z_B = 1.25$~kpc, and normalization
$C_{bulge} = 3.6$.

Since extinction from dust affects IR wavelengths much less than shorter 
wavelengths, this model employs a smooth exponential extinction law.
The assumed absorption is given by:
\be
\frac{d A_\lambda}{ds} = A_{\lambda \circ} 
\exp\left(-{{R-R_\circ}\over{h_{R\,dust}}}
	\,-{{|z|} \over {h_{z\,dust}}}\right),	\label{eq:dust}
\ee
where $A_{\lambda \circ}$ is the passband dependent absorption per unit length 
in 
the solar vicinity, $s$ is the distance along the line of sight to the current 
position, and $h_{R\,dust} = 3.5$~kpc and $h_{z\,dust} = 0.1$~kpc are the 
respective radial and z-component scale lengths for the dust.
The solar neighborhood dust absorption was 
assumed to be 0.07~mag kpc$^{-1}$ at $K$ (\W).  This results in 
$A_{V \circ}=0.62$~mag kpc$^{-1}$ using the interstellar extinction law 
determined by \citet{RL85}.  
This value was simply scaled appropriately for 
other wavelengths.  The extinction calculation adopted the midpoint of the 
line of sight increment as the location of the stellar types.

\subsection{Comparison with Observation}

$K$ predictions of the model have been compared with: the north
galactic pole counts of \citet{El78}; six high galactic latitude
fields of the 2MASS survey; 
% the high galactic latitude survey of \citet[hereafter \G]{GPCM94}; 
and, for completeness, the galactic
plane observations of \citet*[hereafter \E]{EAG84}.  Both \citet{El78}
and \E\ provide cumulative counts while we have computed
differential counts from the 2MASS catalog.

Figure~\ref{fig:elias} shows the comparison of observed cumulative
star counts with model predictions at the NGP.  The model shows
excellent agreement with the \citet{El78} observations.  The model
star counts for each galactic component have also been plotted and
show the dominance of the disk which holds for galactic latitudes $|b|
\geq 20\deg$.  Table~\ref{tab:Eliascomp} gives the over or under
counting percentages versus magnitude.  The model predictions are
within about 20\% of the observed counts.

The partial release of 2MASS data allows us to compare the starcount
model to reality in several high latitude fields.
Figure~\ref{fig:2MASS} shows the comparison between observed
differential counts and model star counts for these regions.  The
model accurately predicts the counts in these regions with the
exception of the low magnitude stars which saturate the 2MASS
detectors.  This saturation may explain the paucity of bright stars
compared to the model.  Table~\ref{tab:2MASScomp} gives the percentage
over or under count in the 2MASS fields versus magnitude.  
The actual counts are on average\footnote{A weighted median was used with
uncertainties given by max$(N^{-1/2},0.1)$.}
$0.89 \pm 0.05$ times
the model counts for the range $6 < K < 12$ in the 4 fields with 
$|b| > 45^\circ$.
This $\sim 10\%$ overprediction in the $6 < K < 12$ range is consistent 
with the $11\%$ overprediction of integrated counts found by \citet{GWC00} in
the $(l,b) = (337^\circ,76^\circ)$ field for the range $9 < K <
12$.

% Figure~\ref{fig:glazebrook} compares six high galactic latitude
% locations of \G\ with differential counts from the model.  Agreement
% of the model with \G\ is fairly good, with the model at least
% following the same trends as the stars with the possible exception of
% the (69\deg,~$-$48\deg) case, where the model increases with
% increasing magnitude while the observed star counts appear to level
% off (The faintest data point of each case is being ignored.  Its
% reliability is shaky since those measurements were made at the
% limiting magnitude of the survey).  According to these observations,
% the model tends to over count the faint end stars.
% %%(THIS IS A DANGEROUS THING TO SAY since overpredicting
% %%stars-->underestimating the CIRB!!!)
% Looking at the model used in Figure~6 of \G\ for comparison, one
% observes a similar excess of faint magnitude stars.  Their model
% consists of only two parts, disk and halo.  The disk is broken up into
% three components: an old stellar disk with a 350~pc scale height; an
% intermediate age disk with a 250~pc scale height; and an old stellar
% extended disk component with a scale height of 1000~pc.  The halo
% involves a disk-shaped luminosity function and a local density scaled
% to 0.0015 that of the disk's.  See \G\ for more details.

One observes an extremely tight correlation between the model and \E's
observations in the galactic plane for all but directly towards the
galactic center and towards $l = 30^\circ$, where the model
respectively over counts brighter stars and under counts overall.
This comparison is shown in Figure~\ref{fig:eaton} and Table~\ref{tab:Ecomp}.
\citet{HCGML99} find the Cohen model under predicts the bright star
contribution near the galactic center when compared with the Two
Micron Galactic plane Survey (TMGS).  The high density of stars in the
galactic center makes confusion a serious problem.  Since at $l =
30$\deg\ one looks tangential to the molecular ring, the under
prediction there probably stems from an incomplete treatment of the
ring component.  More observational insight into the mixture of
stellar types and density of the ring would aid in modeling this
galactic component.  Since we are concerned only with high latitude
regions, we do not pursue either issue further.

We are interested in high galactic latitudes where the contamination
to the CIRB from stars and the ISM is minimal.
Figure~\ref{fig:3obs_diff} synthesizes the model performance for high
galactic latitude regions over a wide range of magnitudes.  We
constructed differential counts from the \citet{El78} results and
computed an average over the six fields of the 2MASS survey.
The model is consistent with the data in these
high latitude regions of interest.  The model is in particularly good
agreement with the recent 2MASS results.
% \\I CAN CONSTRUCT BETTER AVERAGES FROM \G\ and 2MASS DATA

\section{Calculating the CIRB}

\subsection{Model histograms}

The predicted histograms were calculated using a Monte Carlo method based on
$N(<m)$ curves calculated from the \W\ model.
Here $N(<m)$ is the number of stars per unit solid angle brighter than
magnitude $m$.
In order to simulate the DIRBE maps, the flux from a star has to be divided up
among several pixels in the same way that the actual DIRBE divided up the flux.
Flux always extends beyond the 5 pixel blanking area used for bright source
removal by \citet{AOWSH98}, and often extends beyond the 9 pixel blanking area
used by \citet*{KMO96}.
In this work we have divided
the flux into 16 pixels with weights $w_i$ for $i = 1\ldots16$.
The flux division will depend on where within a pixel a star is located, and
will also depend on where the pixel is relative to the cube used to define the
quadrilateralized spherical cube pixel scheme used by COBE.
We have sampled the actual flux division for DIRBE by locating the brightest 8
stars within the region with $\sin|b| > 0.9$.
Thus we obtain 8 different sets of weights, $w_{ij}$, for $i = 1\ldots16$
and $j = 1\ldots8$, which are shown in Table~\ref{tab:beam}.
Note that $\sum_i w_{ij} = 1$ for all $j$.
We divide each DIRBE pixel into 8 subpixels, and use the weights $w_{ij}$
if the star falls into the $j^{th}$ subpixel.

For each sub-pixel one calculates $n$ drawn from a Poisson 
distribution with mean $\mu = 1$.
Then one finds $n$ magnitudes
$m$ chosen so that $\Omega_{sp} N(<m) = U$ where $U$ is a random
number drawn from a distribution uniform between 0 and 1,
and $\Omega_{sp}$ is the solid angle of a sub-pixel.
This can be done quite quickly by making a spline fit to $N(<m)$ 
using $N$ as the independent variable.
The flux of these stars are given by $F = F_0 \, 10^{-0.4m}$, 
where $F_0$ is the flux for a $0^{th}$ magnitude star.
The faintest star produced by this method has magnitude $m_1$ such that
$\Omega_{sp} N(<m_1) = 1$.  
Typical values of $m_1$ are $\approx 11.7$
A Gaussian random number representing stars fainter than $m_1$
is chosen with a mean
\be 
F_{faint} = F_0 \, \Omega_{sp} \int_{m_1}^\infty 10^{-0.4m} [dN(<m)/dm] dm
\ee
and a variance 
\be
\sigma^2_{faint} = F_0^2 \, \Omega_{sp} \int_{m_1}^\infty
10^{-0.8m} [dN(<m)/dm] dm.
\label{eq:sig-faint}
\ee
The flux in each sub-pixel is then weighted by $w_{ij}$, and the resulting
weighted sum of 128 subpixels (8 in each of the 16 neighboring
pixels) gives the Monte Carlo intensity value for one
pixel.  A Gaussian random number with zero mean and standard deviation 
equal to the detector noise given by \citet{HAKDO98} is then added to
the intensity to give the final value.

In order to compute a model histogram this process is repeated many times.
For the model shown in Figure \ref{fig:hist_eye}, nearly 4 million independent
pixel intensity values were generated.  This is 100 times more values than the
39,356 pixels in the region with $\sin|b| > 0.9$ which gives the
data histogram shown in the figure.  
Note that pixel values in the real
maps are not independent, because pixels close enough to be in the DIRBE beam
at the same time are correlated.  Our Monte Carlo method does not model this
correlation because the correlation is a 2-point property of the map, and we
are only comparing histograms which are 1-point properties of the map.

\subsection{Histogram Fitting}
\label{sec:estimate}

To find the CIRB, one simply slides the model 
histogram along the flux axis until it closely matches the DIRBE data
histogram.
The  amount required to align the histograms, $\Delta I$, yields an estimate of
the CIRB.  A ``by eye'' best fit yields 
$\Delta I \approx 12 \; \kJysr$.  
Figure~\ref{fig:hist_eye} shows the 
model histogram along with the DIRBE data histogram on a linear intensity 
scale to facilitate sliding.

We applied a $\chi^2$ method to percentile values extracted from the
histograms to quantify the best fit.  
We found the intensity values for the 5$^{th}$, 15$^{th}$, \ldots 95$^{th}$
\%-tiles in both the model and data histograms.
The model fit assumes that the actual data histogram
intensity at a given percentile, $Y_i$, is given by
$p_1 + p_2 F_i$, where $F_i$ is the corresponding intensity in the model
histogram, $p_1$ is the cosmic background
and $p_2$ is a flux calibration factor.
These percentile values are highly correlated because they are
cumulative statistics.  But the covariance matrix is easily calculated.
Let $I(f)$ be the intensity at the $100f^{th}$ percentile.
The covariance matrix of the percentiles is given by
\be
C_{ij} = \frac{f_i(1-f_j)}{N} 
\left.\frac{\partial I}{\partial f}\right\vert_i
\left.\frac{\partial I}{\partial f}\right\vert_j.
\ee
Given these correlated data points, we minimize $\chi^2$ as follows.
Let $V_{ij} = \partial Y_i/\partial p_j$ be a matrix of partial
derivatives of the data with respect to the parameters.  
Note that $V_{i1} = 1$ and $V_{i2} = F_i$.
Then the parameters are found using
\be
P = [V^T C^{-1} V]^{-1} V^T C^{-1} Y.
\ee
The covariance matrix of the fitted parameters is $[V^T C^{-1} V]^{-1}$.
Because the pixel values in the actual histograms are not independent, we have
scaled the standard deviations of the parameters by $\sqrt{\chi^2_{min}/\nu}$
where $\nu = 8$ is the number of degrees of freedom in the fit.
Figure~\ref{fig:fit} shows the best fit lines for both the two parameter case
and the one parameter fit with $p_2$ forced to be unity.
Once the parameters are found, the model histogram can be scaled and plotted
over the data as shown in Figure~\ref{fig:hist_ks}, which also shows where the
percentiles fall within the distribution.

We have chosen this form of fitting instead of the Kolmogorov-Smirnov test
because the K-S test requires a fixed predefined comparison distribution,
while in our case the model distribution has parameters $p_1$ and $p_2$
which are derived from the data.

\subsection{Tests of the method}

The effect of various perturbations to the modeling on the derived parameters
were tested by simulation.  Three test point spread functions were tried as
replacements for the observed DIRBE beam.  These were a 4 pixel top hat,
and 5 pixel top hat, and beam giving $w_i$ equal to a Gaussian function of $i$
which corresponds to a shape on the sky of $\exp(-\beta\theta^4)$.  The
$w_i$'s for these assumption are listed at the bottom of Table \ref{tab:beam}.
The hard-edged top hat beams lead to a larger value of the CIRB by about
$0.9\;\kJysr$ at 3.5 \um.  The $\exp(-\beta\theta^4)$ beam gave the same CIRB
as the observed DIRBE beam to within $0.1\;\kJysr$.
We have also run eight tests using just one of the star profiles at a time.
The standard deviation of the mean of these tests is 0.11~\kJysr\ at 
2.2 \um\ and 0.06~\kJysr\ at 3.5 \um.

Even though our method is independent of the relative calibration between
DIRBE fluxes and magnitudes, a comparison with previous work requires knowledge
of $F_0$.  
While most of the stars in Table \ref{tab:beam} are variable, we can 
still find their median 3.5 \um\ magnitudes in \citet*{GSM87} and compute 
values for $F_0(L)$ at 3.5 \um\ which are given in Table~\ref{tab:Lcal}.  
The median of these $F_0(L)$'s is 254~Jy and the mean is $265\pm12$~Jy which 
agree with the $F_0(L) = 263$~Jy in \citet{GWC00}.

Tests that excluded the detector noise ($\sigma = 1.05\;\kJysr$ per pixel
at 3.5 \um)
given by \citet{HAKDO98} changed the derived CIRB by 
$-0.20\;\kJysr$ at 2.2 \um\ and
$-0.30\;\kJysr$ at 3.5 \um.  The noise quoted by \citet{HAKDO98} may contain
some confusion noise due to unresolved stars which are already included in 
our Monte Carlo histograms and should not be counted twice.

The histogram fitting method is independent of the adopted 
DIRBE flux at $0^{th}$ 
magnitude because the free parameter $p_2$ multiplies the model fluxes.
Since the star counts in the region of interest are nearly a power law,
this also makes the derived CIRB values nearly independent of a constant
scaling factor applied to the counts, since for a power law, scaling the counts
is equivalent to a scaling of the fluxes.  We have adopted a count scaling
factor of 0.9 based on comparison of the model to the high latitude 2MASS
fields, and varying this factor by $\pm 0.1$ only changes the derived CIRB by 
$\mp 0.52\;\kJysr$ at 2.2 \um\ and $\mp 0.43\;\kJysr$ at 3.5 \um.

Any change in the differential source counts, $dN/dm$, from the starcount
model will have some effect on the derived CIRB, and different changes to the
model that produce the same change in $dN/dm$ will produce the same change in
the CIRB.  Table~\ref{tab:bump} shows how the derived CIRB changes when
the differential star counts are increased by a ``bump'' that is 25\% high,
centered at $m_b$ and 1 magnitude wide (FWHM):
\be
\left(\frac{dN}{dM}\right)^\prime = \exp(0.25\max(0,1-|m-m_b|))\frac{dN}{dM}.
\ee
Increasing the bright star count causes the derived CIRB to go up, while
increasing the faint star count causes the derived CIRB to go down.
The corresponding changes in the derived CIRB using the \citet{AOWSH98}
method, shown in the last two columns of Table~\ref{tab:bump},
are generally larger and always cause the CIRB to go down.

The model has an outer radius of 15~kpc.  Changing this cutoff to 25~kpc
reduces the derived CIRB by only $-0.08\;\kJysr$ at 2.2 \um\ 
and $-0.13\;\kJysr$ at 3.5 \um.  
If we triple the halo density of all lower main sequence stars
(later than G2V) this changes the derived CIRB 
by only $-0.58\;\kJysr$ at 2.2 \um\ and $-0.20\;\kJysr$ at 3.5 \um\ in the
HQB region.
These results are not surprising since the halo makes such a small 
contribution to the star counts in Figure~\ref{fig:elias}.

We also used several regions to test for isotropy:
the NGP with $\sin(b) > 0.96$, the
SGP with $\sin(b) < -0.96$, the NEP with $\sin\beta > 0.96$, 
a region with $|\sin|b|-\sqrt{1/2}| < 0.035$ and $|\beta| > 45^\circ$ (B45),
and the High Quality B (HQB) region from \citet{HAKDO98}
with $|b| > 60^\circ$ and $|\beta| > 45^\circ$.  
The CIRB's derived from these regions, shown
in Table \ref{tab:isotropy}, are very consistent with each other and with the
values derived by \citet{GWC00} by direct measurement and
subtraction of stars in the ``dark spot'' in the 2.2 \um\ band, and
reasonably consistent at 3.5 \um.  If we fit the values in 
Table~\ref{tab:isotropy} to $A+G\csc|b|$ or $A+Z\csc|\beta|$, 
we find no trend with either $b$ or $\beta$ at 2.2 \um, but a slope of 
$Z = 2.65\;\kJysr$ per unit of $\csc|\beta|$ at 3.5 \um.  
This indicates a problem
with our zodiacal light model but the slope is smaller than our zodiacal 
model error estimate of 3.3~\kJysr.

Finally, in regions with a gradient in the star density, such as the NEP
region, it is important to allow for this gradient.   Making an average
starcount by running the starcount model for several subregions, and
then generating one histogram for the  whole region from the average
starcounts, will give an incorrect result for the CIRB.  The histogram
making operation and the averaging operation do not commute, and the
pixel values that go into the actual data histogram are generated from
the stars within a single instrument field of view. The correct
procedure is to calculate the starcount model for many small subregions,
generate individual histograms for these subregions, and then average
the histograms together.  We have used between 4 and 8 subregions in our 
fields.

The position of the Sun near the edge of a spiral arm leads to a 
discontinuity across the sky on a line through the galactic poles from 
longitude $l = \tan^{-1}(4.57) = 78^\circ$ to $l = 258^\circ$.
The amplitude of
this discontinuity is 9\% in $dN/dm$ at $L = 7$ and 4.6\% in the total
integrated intensity at 3.5 \um.  Analyzing the NGP data using all subregions,
both inside and outside the discontinuity, gives a CIRB of 16.65 
and 14.27~\kJysr\ at 2.2 and 3.5 \um, as reported in Table~\ref{tab:isotropy}.
Using model histograms computed only from star counts inside the discontinuity
gives a CIRB of 16.16 and 14.07~\kJysr, while using only star counts from
outside the discontinuity gives 16.39 and 14.51~\kJysr.  Thus while the model
is not perfect, the histogram method is quite insensitive to its faults.

\section{Discussion}

At this point we have independent estimates for the CIRB at 2.2 and 3.5 \um.
These estimates from the histogram fitting method are consistent with the
independent estimates of the CIRB obtained by direct subtraction of measured
stars in \citet{GWC00}.  Now we can combine the 2.2 and 3.5 \um\ maps
following the technique of \citet{DA98}, which gives a CIRB estimate at 3.5
\um\ if the 2.2 \um\ CIRB is known.

We proceed by fitting the 3.5 \um\ map to 
a linear combination of the 2.2 \um\ map, the 100 \um\ map, a constant and
$\csc\beta$.  A fit that minimzes the sum of absolute values of the errors
in the region $\sin|b| > 0.4$ and $\sin|\beta| > 0.4$ gives
\be
I_{3.5} = 2.68 + 2.66 \csc|\beta| + 0.4976 I_{2.2} + 0.00094 I_{100}
\;\kJysr.
\label{eq:RLLAV}
\ee
This fit gives the same slope \vs\ $\csc|\beta|$ as the fit to the data in
Table~\ref{tab:isotropy}.  The coefficient of the 2.2 \um\ map becomes
0.3128 when the fit is expressed in \nWmmsr\ instead of \kJysr, in excellent
agreement with the fit found by \citet{DA98}.  The coefficient of
the 100 \um\ map is only one-half of the value found by \citet{AOWSH98}
in regions close to the galactic plane.  The constant term 
of 2.68~\kJysr\ is not the CIRB, but rather the displacement of the 
3.5 \um\ \vs\ 2.2 \um\ correlation from the origin (see eq.~[\ref{eq:R35}]
for the CIRB).
Repeating the fit in equation~\ref{eq:RLLAV} using the DIRBE ZSMA
maps produced using the \citet{KWFRA98}
model gives $3.57 \csc|\beta|$ for the dependence on ecliptic latitude, so
our new zodiacal light model gives a smaller residual slope than the
\citet{KWFRA98} model.  Note that while a non-zero slope \vs\ $\csc|\beta|$
clearly indicates a problem with the zodiacal model, a small slope does not
necessarily mean a small zero-point error.

We can construct a map of the 3.5 \um\ residual after subtracting the 
galactic contributions due to stars and the ISM using
\bea
R_{3.5} & = & I_{3.5} - 0.4976 (I_{2.2} - I_\circ(2.2)) - 0.00094 (I_{100}
- I_\circ(100)) \nonumber \\
& = & I_{3.5} - 0.4976 I_{2.2} - 0.00094 I_{100} +
8.8\;\kJysr
\label{eq:R35}
\eea
where $I_\circ(2.2) = 16.9\;\kJysr$ and $I_\circ(100) = 0.4\;\MJysr$.
Figure~\ref{fig:L-K-100} shows a map of this residual which is nearly
isotropic over the high galactic latitude sky.  Figure~\ref{fig:hist-2}
compares the histogram of the original zodi-subtracted 3.5 \um\ map
in the region with $b > 45^\circ$ and $\beta > 45^\circ$ to the histogram
of the residual map in same region.  The histogram of the residual,
shown in Figure~\ref{fig:hist-2}, is very
sharply peaked.  A Gaussian fit to the highest three bins gives
$14.23 \pm 1.57\;\kJysr$ for the mean residual and its single pixel 
standard deviation.  Both the map and the histogram show that the
CIRB estimate obtained by combining the 2.2 and 3.5 \um\ maps is quite
isotropic.

Based on our tests, we have set up the error budget shown in 
Table~\ref{tab:budget}.  The largest term is the zodiacal light error
which is estimated at 5\% of the zodiacal intensity at the ecliptic poles.
The next largest term is an estimate for errors from the starcount model 
other than constant count scaling or flux scaling factors, for which we have
used the quadrature sum of the values in Table~\ref{tab:bump}, which were
based on 25\% errors in the starcounts.
The ISM error is
taken as 50\% of the ISM correction in the HQB region.
The DIRBE noise error is taken as 50\% of the difference between the no noise
and with noise CIRB's.
The DIRBE beam error is based on the scatter among the eight different stars
in Table~\ref{tab:beam} when used one at a time.  
The standard deviation of the mean of the five histogram fitting results 
in Table~\ref{tab:isotropy} is included as ``scatter''.
The quadrature sum of the errors is used as our final uncertainty estimate.

Given that three independent methods show the existence of a CIRB at 2.2 and
3.5 \um, it is worth investigating why \citet{HAKDO98} did not find it.
One reason is that the mean of the unblanked pixels, which was used by
\citet{AOWSH98}, is a very inefficient statistical estimator when the noise
is dominated by confusion.  Consider $N$ samples $x_i = C + y_i$,
where $y_i$ are independent identically distributed random variables
with known probability density function $p(y)$.  Here the $x_i$ are pixel data
values, $C$ is the CIRB, and $p(y)$ is the normalized model histogram.  Then
the likelihood $L$ is given by
\bea
\ln[L(\hat{C})] & = & \sum_{i=1}^N \ln[p(x_i-\hat{C})]
\nonumber\\
& \approx & N \int \ln[p(x-\hat{C})] p(x-C) dx
\nonumber\\
& \approx & \ln[L_{max}] - \frac{N}{2\sigma_1^2}(\hat{C}-C)^2 +\ldots
\eea
Using the model histogram in Figure~\ref{fig:hist_eye}, we find that
standard deviation of the CIRB for a sample size of 1 pixel is
$\sigma_1 = 6.5\;\kJysr$.  But the corresponding standard deviation for
the mean of the unblanked pixels is given by 
$\sigma_{faint} = 50.3\;\kJysr$ from
equation (\ref{eq:sig-faint}) with the one pixel $\Omega$ and $m_1$ set to the
blanking level of $L = 3$.  Thus the histogram fitting method would reach
a statistical noise level of $1\;\kJysr$ for $N = 43$ pixels while the
mean of unblanked pixels needs 2,526 pixels to reach the same noise level.

However, since DIRBE observed $>10^5$ high latitude pixels, statistical
efficiency is not a major concern, but systematic error sensitivity is.
The faint source model (FSM) of \citet{AOWSH98} is the same as the model we 
have used to predict model histograms, but \citet{AOWSH98} were much more 
sensitive to model parameters than this paper.  
The FSM in \citet{AOWSH98} contributed 67~\kJysr\ per unit of $\csc|b|$ 
at 2.2 \um\ and 40~\kJysr\ at 3.5 \um.  
Thus the sensitivity of the CIRB to 
$\pm 10\%$ model over or under predictions  at 2.2 \um\ is
about $\mp 6.7\;\kJysr$ which is 12.8 times higher than the 
sensitivity of our histogram method.
At 3.5 \um\ the \citet{AOWSH98} method is 9.2 times
more sensitive to model over or under predictions than our histogram method.  
And the histogram method is completely insensitive to changes in the flux 
at 0$^{th}$ magnitude while the \citet{AOWSH98} method is directly
proportional to errors in $F_0$.
% Finally, there is a problem at the boundary between the bright source
% blanking and the faint source subtraction in the \citet{AOWSH98} method. 
% The blanking threshold was 15 Jy at 2.2 and 3.5 \um, but a source just
% below the blanking threshold, say a 12 Jy source, is quite likely to
% have a nearby 10 Jy source within a $3\times3$ pixel box, which will
% push the combined flux above the blanking threshold.   Thus confusion
% noise will cause some sources just below the blanking  threshold to be
% blanked, and these sources are both blanked and subtracted,  leading to
% a systematic oversubtraction by the FSM.
The combined effects of 
% this systematic oversubtraction,
a 10\% overprediction by the model, and an 8\% change in $F_0$
at 3.5 \um\ \citep{GWC00}, 
explain most of the galactic slope in the residual maps of \citet{AOWSH98}
which was -27\% of the FSM slope.
At 2.2 \um, \citet{GWC00} found the \citet{AOWSH98} $F_0$ was correct,
so one would expect the galactic slope in the residual maps 
of \citet{AOWSH98} to be reduced to -19\%,
while it was actually -18\% of the FSM.

\section{Conclusion}
\label{sec:conclusion}

Using the unweighted mean of the 5 histogram fitting results
in Table~\ref{tab:isotropy},
we obtain an estimate of the CIRB 
of \KkJy\ at 2.2 \um\ and \LkJy\ at 3.5 \um, 
where the errors are dominated
by zodiacal light model uncertainties.
These values compare quite well with the values obtained by \citet{GWC00}
using direct subtraction of measured stars in the dark spot. 
Thus we have obtained CIRB values at 2.2 and 3.5 \um\ that are consistent 
with the dark spot values, even though the sky areas used are disjoint and 
the techniques used are very different.  
Our values are also consistent with the \citet{HAKDO98}
upper limits, and with the \citet{DA98} correlation between the
2.2 and 3.5 \um\ CIRB values, as shown in Table~\ref{tab:CIRBall}.
The dominant uncertainty is in the zodiacal light
model, which gives a systematic error common to both the dark spot and 
the histogram values for the CIRB.

Our values show that the near IR background has a bolometric intensity that
is similar to the bolometric intensity of the far IR background found
by \citet{HAKDO98}.  Thus roughly 50\% of the radiation produced by
galaxies is absorbed by dust and re-radiated in the far IR.

\acknowledgments

EDR acknowledges support from NASA GSRP Fellowship NGT5-50173.
The {\sl COBE} datasets were developed by the NASA Goddard Space Flight
Center under the guidance of the COBE Science Working Group and were
provided by the NSSDC.  This publication makes use of data products from the
Two Micron All Sky Survey, which is a joint project of the University of
Massachusetts and the Infrared Processing and Analysis Center, funded by the
National Aeronautics and Space Administration and the National Science
Foundation.  We thank Rick Arendt for making many useful comments and 
criticisms of a draft of this paper.

\clearpage

%%%%%%%%%%%%%%%%%%%%%%%		Tables
% order of reference in paper:
% \ref{tab:spiral}
% \ref{tab:Eliascomp}
% \ref{tab:2MASScomp}
% \ref{tab:Ecomp}
% \ref{tab:beam}
% \ref{tab:Lcal}
% \ref{tab:bump}
% \ref{tab:isotropy}
% \ref{tab:budget}
% \ref{tab:CIRBall}

\begin{deluxetable}{cccccc}
\singlespace
%\footnotesize
\tablecaption{Spiral Arm Representation \label{tab:spiral}}
\tablehead{
\colhead{Arm/Spur} & \colhead{$\alpha$} & \colhead{$R_{min}$} & 
\colhead{$\theta_{min}$} & \colhead{$E$} & \colhead{$W$} \\
\colhead{index}&& \colhead{(kpc)} & \colhead{(rad)} & \colhead{(rad)} & 
\colhead{(kpc)}
}
\tablewidth{0pt}
\tablecolumns{6}
\startdata
1 & 4.25 & 3.48 & 0.000 & 6.00 & 0.75 \\
2 & 4.25 & 3.48 & 3.141 & 6.00 & 0.75 \\
3 & 4.89 & 4.90 & 2.525 & 6.00 & 0.75 \\
4 & 4.89 & 4.90 & 5.666 & 6.00 & 0.75 \\
\tableline
5 & 4.57 & 8.10 & 5.847 & 0.55 & 0.30 \\
6 & 4.57 & 7.59 & 5.847 & 0.55 & 0.30 \\
\enddata
%\tablenotetext{a}
%\tablerefs{}
%\tablecomments{}
\end{deluxetable}

\begin{deluxetable}{rr}
\tablecaption{$100\times\ln(\mbox{ACTUAL/MODEL})$ for 
the NGP region (Elias)\label{tab:Eliascomp}}
\tablewidth{0pt}
\tablehead{
\colhead{mag}    & 
\colhead{$b=90$}
}
\startdata
 1.00 &$ -18 \pm 39 $ \\
 2.00 &$ -54 \pm 30 $ \\
 3.00 &$ -11 \pm 21 $ \\
 3.50 &$ -22 \pm 17 $ \\
 6.50 &$  34 \pm 42 $ \\
 7.50 &$ -10 \pm 33 $ \\
 8.50 &$   5 \pm 24 $ \\
\enddata
\end{deluxetable}

\begin{deluxetable}{rrrrrrr}
\tablecaption{$100\times\ln(\mbox{ACTUAL/MODEL})$ counts in
regions from the 2MASS catalog.\label{tab:2MASScomp}}
\tablewidth{0pt}
\tablehead{
 &
\colhead{$l=34$}  &
\colhead{$l=286$} &
\colhead{$l=341$} &
\colhead{$l=225$} &
\colhead{$l=205$} &
\colhead{$l=158$} \\
 &
\colhead{$b=82$} &
\colhead{$b=77$} &
\colhead{$b=75$} &
\colhead{$b=54$} &
\colhead{$b=19$} &
\colhead{$b=-41$} \\
\colhead{mag} & 
\colhead{$\Omega = 7.9(^\circ)^2$}  & 
\colhead{$\Omega = \pi(^\circ)^2$}  &
\colhead{$\Omega = 29.5(^\circ)^2$} &
\colhead{$\Omega = \pi(^\circ)^2$}  &
\colhead{$\Omega = \pi(^\circ)^2$}  &
\colhead{$\Omega = 3.9(^\circ)^2$}
}
\startdata
 3.5 &     \nodata &     \nodata &     \nodata &     \nodata &     \nodata &     \nodata \\
 4.5 &     \nodata &     \nodata & $-100\pm58$ &     \nodata &     \nodata & $  27\pm71$ \\
 5.5 & $ -89\pm71$ &     \nodata & $ -54\pm31$ & $ -76\pm99$ & $  61\pm34$ & $  51\pm41$ \\
 6.5 & $  -4\pm32$ & $  -3\pm50$ & $ -24\pm18$ & $ -43\pm58$ & $   3\pm29$ & $ -49\pm45$ \\
 7.5 & $ -51\pm29$ & $ -46\pm45$ & $ -29\pm14$ & $ -41\pm41$ & $ -12\pm21$ & $  -5\pm25$ \\
 8.5 & $  -2\pm17$ & $  -4\pm27$ & $   4\pm09$ & $ -41\pm31$ & $ -49\pm17$ & $   0\pm18$ \\
 9.5 & $   8\pm12$ & $ -33\pm22$ & $   5\pm06$ & $ -21\pm20$ & $  -9\pm10$ & $  13\pm13$ \\
10.5 & $   1\pm08$ & $  10\pm12$ & $ -13\pm05$ & $ -11\pm13$ & $  12\pm07$ & $  -8\pm10$ \\
11.5 & $ -13\pm06$ & $ -13\pm09$ & $ -16\pm03$ & $ -10\pm08$ & $   8\pm05$ & $   6\pm06$ \\
12.5 & $ -22\pm04$ & $ -19\pm07$ & $ -19\pm02$ & $ -12\pm06$ & $  10\pm04$ & $ -11\pm05$ \\
13.5 & $  -8\pm03$ & $  -8\pm05$ & $  -6\pm02$ & $   2\pm04$ & $   6\pm03$ & $ -22\pm04$ \\
14.5 & $  30\pm02$ & $  26\pm04$ & $  32\pm01$ & $  21\pm03$ & $   9\pm02$ & $  15\pm03$ \\
15.5 & $  19\pm02$ & $  62\pm03$ & $  36\pm01$ & $  36\pm03$ & $ -24\pm02$ & $ -47\pm03$ \\
\enddata
\end{deluxetable}

\begin{deluxetable}{rrrrrrrr}
\tablecaption{$100\times\ln(\mbox{ACTUAL/MODEL})$ for 
the \E\ regions\label{tab:Ecomp}}
\tabletypesize{\footnotesize}
\tablewidth{0pt}
\tablehead{
\colhead{mag} & 
\colhead{$l=\phn 0$}  & 
\colhead{$l=10$} & 
\colhead{$l=20$} & 
\colhead{$l=30$} & 
\colhead{$l=40$} &
\colhead{$l=50$} &
\colhead{$l=60$} 
\\
%\colhead{mag}    & 
%\colhead{$b=$} & 
%\colhead{$b=$} &
%\colhead{$b=$} & 
%\colhead{$b=$} & 
%\colhead{$b=$} & 
%\colhead{$b=$} &
%\colhead{$b=$} \\
\colhead{} &
\colhead{$\Omega=144\, (\arcmin)^2$} &
\colhead{$\Omega= 99\, (\arcmin)^2$} &
\colhead{$\Omega=103\, (\arcmin)^2$} &
\colhead{$\Omega=384\, (\arcmin)^2$} &
\colhead{$\Omega=104\, (\arcmin)^2$} &
\colhead{$\Omega=121\, (\arcmin)^2$} &
\colhead{$\Omega=101\, (\arcmin)^2$} 
}
\startdata
 5.00 &    \nodata      &    \nodata   &    \nodata   &$  -7 \pm    58 $&    \nodata      &    \nodata      &   \nodata      \\
 5.50 &    \nodata      &    \nodata   &    \nodata   &$  49 \pm    33 $&    \nodata      &    \nodata      &   \nodata      \\
 6.00 &    \nodata      &    \nodata   &    \nodata   &$  58 \pm    24 $&    \nodata      &    \nodata      &   \nodata      \\
 6.50 &    \nodata      &    \nodata   &    \nodata   &$  43 \pm    20 $&    \nodata      &    \nodata      &   \nodata      \\
 7.00 &$-216 \pm    58 $&$ -29 \pm 45 $&$ -26 \pm 41 $&$  43 \pm    16 $&$ -83 \pm    71 $&$ -43 \pm    58 $&$  -5 \pm    58 $\\
 7.50 &$-244 \pm    41 $&$  -2 \pm 30 $&$  10 \pm 27 $&$  33 \pm    13 $&$ -21 \pm    41 $&$ -40 \pm    45 $&$  -2 \pm    45 $\\
 8.00 &$-178 \pm    21 $&$ -17 \pm 26 $&$   5 \pm 22 $&$  36 \pm    11 $&$  -7 \pm    30 $&$ -37 \pm    35 $&$   1 \pm    35 $\\
 8.50 &$-110 \pm    11 $&$ -18 \pm 20 $&$  18 \pm 17 $&$  32 \pm\phn 9 $&$  13 \pm    22 $&$  -5 \pm    24 $&$   6 \pm    28 $\\
 9.00 &$ -24 \pm\phn 7 $&$   4 \pm 15 $&$  18 \pm 14 $&$  26 \pm\phn 7 $&$  -5 \pm    19 $&$   0 \pm    19 $&$  -1 \pm    23 $\\
 9.50 &$   9 \pm\phn 5 $&$  12 \pm 12 $&$  19 \pm 11 $&$  23 \pm\phn 7 $&$ -11 \pm    16 $&$   1 \pm    15 $&$  -2 \pm    19 $\\
10.00 &$ -19 \pm\phn 5 $&$  -3 \pm 11 $&$  26 \pm 10 $&$   9 \pm\phn 6 $&$ -15 \pm    13 $&$  -3 \pm    13 $&$ -20 \pm    16 $\\
10.50 &    \nodata      &$ -28 \pm 10 $&$ 25 \pm\phn 9 $&$ 7 \pm\phn 5 $&$  -8 \pm    11 $&$   5 \pm    10 $&$  -7 \pm    13 $\\
11.00 &    \nodata      &    \nodata   &    \nodata   &   \nodata 	&$  11 \pm\phn 9 $&$   6 \pm\phn 8 $&$   9 \pm    10 $\\
11.50 &    \nodata      &    \nodata   &    \nodata   &   \nodata	&$   5 \pm\phn 8 $&$   7 \pm\phn 7 $&$   3 \pm\phn 8 $\\
\enddata
\end{deluxetable}

\begin{deluxetable}{lrrrrrrrrrrrrrrrr}
\singlespace
\tabletypesize{\scriptsize}
% \rotate
\tablecaption{DIRBE Beam.  $10^4$ times the flux fraction in pixels
from 1 at the peak to the 16$^{th}$ nearest pixel\label{tab:beam}}
\tablewidth{0pt}
\tablehead{
\colhead{Source} &
\colhead{1} & \colhead{2} & \colhead{3} & \colhead{4} & \colhead{5} & 
\colhead{6} & \colhead{7} & \colhead{8} & \colhead{9} & \colhead{10} & 
\colhead{11} & \colhead{12} & \colhead{13} & \colhead{14} & \colhead{15} & 
\colhead{16} 
}
\tablecolumns{17}
\startdata
$\alpha$~Boo & 2495 & 1724 & 1641 & 1209 & 1286 &  683 &  635 &  105 &  181 &    8 &   13 &    7 &    3 &    3 &    4 &    2 \\
RX Boo       & 2389 & 2251 &  917 & 1460 &  392 &   98 & 1641 &   67 &  164 &  327 &    4 &   55 &  217 &    5 &    4 &    7 \\
R Aqr        & 2314 & 2134 &  750 & 1373 & 1141 &   81 & 1052 &  709 &  197 &  118 &   45 &   13 &    8 &   25 &   27 &   12 \\
RT Vir       & 2278 & 1839 &  669 &  670 & 2125 &  459 &  184 & 1521 &    9 &   38 &    8 &   11 &  166 &    7 &   11 &    7 \\
$\delta$~Vir & 2274 &  322 & 2149 & 2258 &  302 &  184 &  241 &   13 & 2000 &   20 &  119 &   68 &   11 &   11 &   16 &   13 \\
Y CVn        & 2207 &  645 & 2088 & 1650 &   52 & 1788 &    8 &  240 &    8 &    8 &  646 &  625 &    9 &    9 &    9 &    7 \\
BK Vir       & 2244 &  264 & 2243 &  591 & 1769 & 1804 &   18 &   96 &  519 &   14 &  227 &   15 &   49 &  105 &   22 &   21 \\
T Cet        & 2206 & 1968 &  861 & 1999 &  321 &   10 & 1618 &  154 &  619 &   68 &   15 &   72 &   31 &   16 &   29 &   14 \\
\tableline
Top Hat 4    & 2500 & 2500 & 2500 & 2500 &    0 &    0 &    0 &    0 &    0 &    0 &    0 &    0 &    0 &    0 &    0 &    0 \\
Top Hat 5    & 2000 & 2000 & 2000 & 2000 & 2000 &    0 &    0 &    0 &    0 &    0 &    0 &    0 &    0 &    0 &    0 &    0 \\
$e^{-\beta \theta^4}$ & 2253 & 2033 & 1713 & 1348 &  990 &  680 &  435 &  261 &  145 &   76 &   37 &   17 &    7 &    3 &    1 &    0 \\
\enddata
\end{deluxetable}

\begin{deluxetable}{lrrrr}
\tablecaption{DIRBE Calibration at 3.5 \um.\label{tab:Lcal}}
\tablewidth{0pt}
\tablehead{
\colhead{Source} & \colhead{\#\tablenotemark{a}} & \colhead{median L} & 
\colhead{DIRBE Flux [Jy]} & \colhead{$F_0$ in Jy} 
}
\startdata
$\alpha$~Boo & 23 & -3.12 &  4500 & 254.2 \\
RX Boo       & 10 & -2.30 &  2009 & 241.5 \\
R Aqr        &  4 & -1.50 &  1327 & 334.9 \\
RT Vir       &  3 & -1.40 &  1073 & 295.5 \\
$\delta$~Vir &  6 & -1.39 &   915 & 254.3 \\
Y CVn        &  8 & -1.51 &   913 & 227.2 \\
BK Vir       &  2 & -1.26 &   805 & 252.2 \\
T Cet        &  0 & -1.14\tablenotemark{b} &   753 & 262.9 \\
\enddata
\tablenotetext{a}{The number of magnitudes for $3.4 \leq \lambda \leq
3.6\;\um$ in \protect\citet{GSM87}.}
\tablenotetext{b}{Interpolated between the 2.2 and 4.2 \um\ magnitudes.}
\end{deluxetable}

\begin{deluxetable}{rrrrr}
\tablecaption{Effect on the CIRB of increasing the differential star
counts by 25\% over a 1 mag FWHM bin centered at $m_b$.\label{tab:bump}}
\tablehead{
& \multicolumn{2}{c}{This paper} 
& \multicolumn{2}{c}{\protect\citet{AOWSH98}} \\
\colhead{$m_b$}
& \colhead{$\Delta I_{CIRB}(2.2\;\um)$}
& \colhead{$\Delta I_{CIRB}(3.5\;\um)$}
& \colhead{$-\Delta I(K>4)$}
& \colhead{$-\Delta I(L>3)$}
}
\tablewidth{0pt}
\startdata
 2 & $0.45\;\kJysr$  & $0.07\;\kJysr$  & $-0.00\;\kJysr$ & $-0.00\;\kJysr$ \\
 3 & $0.82\;\kJysr$  & $0.16\;\kJysr$  & $-0.00\;\kJysr$ & $-0.66\;\kJysr$ \\
 4 & $0.90\;\kJysr$  & $0.40\;\kJysr$  & $-1.13\;\kJysr$ & $-1.10\;\kJysr$ \\
 5 & $0.96\;\kJysr$  & $0.38\;\kJysr$  & $-2.19\;\kJysr$ & $-1.04\;\kJysr$ \\
 6 & $0.23\;\kJysr$  & $-0.08\;\kJysr$ & $-1.99\;\kJysr$ & $-0.94\;\kJysr$ \\
 7 & $-0.39\;\kJysr$ & $-0.50\;\kJysr$ & $-1.66\;\kJysr$ & $-0.77\;\kJysr$ \\
 8 & $-0.65\;\kJysr$ & $-0.55\;\kJysr$ & $-1.25\;\kJysr$ & $-0.58\;\kJysr$ \\
 9 & $-0.45\;\kJysr$ & $-0.23\;\kJysr$ & $-0.92\;\kJysr$ & $-0.44\;\kJysr$ \\
10 & $-0.63\;\kJysr$ & $-0.29\;\kJysr$ & $-0.76\;\kJysr$ & $-0.37\;\kJysr$ \\
11 & $-0.44\;\kJysr$ & $-0.33\;\kJysr$ & $-0.71\;\kJysr$ & $-0.34\;\kJysr$ \\
12 & $-0.31\;\kJysr$ & $-0.40\;\kJysr$ & $-0.63\;\kJysr$ & $-0.30\;\kJysr$ \\
\enddata
\end{deluxetable}

\begin{deluxetable}{lrrrrr}
\tablecaption{Comparison of Different Regions: Statistical Errors Only
\label{tab:isotropy}}
\tablehead{
&&&&\colhead{\kJysr\ at}&\colhead{\kJysr\ at}\\
\colhead{Region} & \colhead{\# of pixels}
& \colhead{$\langle \csc|b| \rangle$} & \colhead{$\langle \csc|\beta| \rangle$}
& \colhead{2.2 \um} & \colhead{3.5 \um}
} 
\tablewidth{0pt}
\tablecolumns{6}
\startdata
Dark Spot   &   17 & 1.095 & 1.289 & $16.4 \pm 2.3$ & $12.8 \pm 1.8$ \\
% NGP       & 7875 & 1.021 & 2.204 & $14.9 \pm 0.4$ & $14.6 \pm 0.2$ \\
% SGP       & 7875 & 1.021 & 2.204 & $16.2 \pm 0.5$ & $17.0 \pm 0.2$ \\
% HQB       & 8688 & 1.100 & 1.296 & $15.3 \pm 0.5$ & $13.9 \pm 0.2$ \\
% B45       & 8372 & 1.415 & 1.151 & $15.5 \pm 0.4$ & $12.9 \pm 0.2$ \\
% NEP       & 7816 & 2.204 & 1.021 & $15.5 \pm 1.0$ & $11.6 \pm 0.6$ \\
% Mean = 15.6+/-0.6(sd1); 13.8+/-1.9(sd1)
% \tableline
% with floating flux cal
NGP       & 7875 & 1.021 & 2.204 & $16.7 \pm 0.4$ & $15.0 \pm 0.3$ \\
SGP       & 7875 & 1.021 & 2.204 & $18.9 \pm 0.5$ & $17.7 \pm 0.3$ \\
HQB       & 8688 & 1.100 & 1.296 & $16.9 \pm 0.9$ & $14.2 \pm 0.4$ \\
B45       & 8372 & 1.415 & 1.151 & $17.3 \pm 0.6$ & $13.4 \pm 0.4$ \\
NEP       & 7816 & 2.204 & 1.021 & $14.9 \pm 2.0$ & $11.9 \pm 1.2$ \\
% median K=16.8, mean = 16.85 +/- 0.53; mean of this paper = 16.94
% median L=13.8, mean = 14.17 +/- 0.83; mean of this paper = 14.44
\enddata
\end{deluxetable}

\begin{deluxetable}{lrrr}
\tablecaption{CIRB error budget.\label{tab:budget}}
\tablehead{
\colhead{Term} & \colhead{Uncertainty} & \colhead{$\Delta I_{CIRB}(2.2\um)$} 
& \colhead{$\Delta I_{CIRB}(3.5\um)$} }
\tablewidth{0pt}
\startdata
Zodiacal	&   5\% & 3.79~\kJysr &  3.25~\kJysr \\
ISM             &  50\% & \nodata     &  0.85~\kJysr \\
Starcount       &  25\% & 2.03~\kJysr &  1.14~\kJysr \\
Count scaling	&   5\% & 0.26~\kJysr &  0.22~\kJysr \\
Flux scaling	&   5\% & \nodata     &  \nodata     \\
DIRBE noise     &  50\% & 0.10~\kJysr &  0.15~\kJysr \\
DIRBE beam      & 100\% & 0.11~\kJysr &  0.06~\kJysr \\
scatter         & 100\% & 0.64~\kJysr &  0.96~\kJysr \\
\tableline
Quadrature Sum  &       & 4.36~\kJysr &  3.69~\kJysr \\
\enddata
\end{deluxetable}

\begin{deluxetable}{rll}
\singlespace
%\footnotesize
\tablecaption{CIRB at 3.5 $\mu$m \label{tab:CIRBall}}
\tablehead{
\colhead{flux} & \colhead{detector} & \colhead{reference} 
}
\tablewidth{0pt}
\tablecolumns{3}
\startdata
 $     35.00\;\kJysr$     & theory & \citet{PP67}\\
 $    128.56\;\kJysr$     & rocket & \citet*{MAM88}\\
% MAM88 give (4.8+/-1.7)E-12 W/cm^2/um/sr = 77.44 kJy/sr at 2.2
% (2.9+/-0.9)E-12 W/cm^2/um/sr = 139.59 kJy/sr at 3.8
% 3.2+/-1.3)E-12 W/cm^2/um/sr = 131.41 kJy/sr at 3.51
 $    <271.83\;\kJysr$     & rocket & \citet{MKMMN94}\\
% CIBR < 200 nW/m^2/sr at 2.5 um, < 250 nW/m^2/sr at 4 um
 $    <26.83\;\kJysr$     & DIRBE  & \citet{HAKDO98}\\
 $    >11.55\;\kJysr$     & DIRBE  & \citet{DA98}\\
 $   \LGWCkJyval\;\kJysr$ & DIRBE  & \citet{GWC00}\\
 $   \LkJyval\;\kJysr$     & DIRBE  & this paper\\
\enddata
%\tablenotetext{a}
%\tablerefs{}
%\tablecomments{}
\end{deluxetable}

\clearpage

%%%%%%%%%%%%%%%%%%%%%%		Figures
% order of reference in paper:
% \ref{fig:spiral}
% \ref{fig:elias}
% \ref{fig:2MASS}
% \ref{fig:eaton}
% \ref{fig:3obs_diff}
% \ref{fig:hist_eye}
%  need to refer to fit
%  need to refer to hist_ks
% \ref{fig:L-K-100}
% \ref{fig:hist-2}

\myfigure{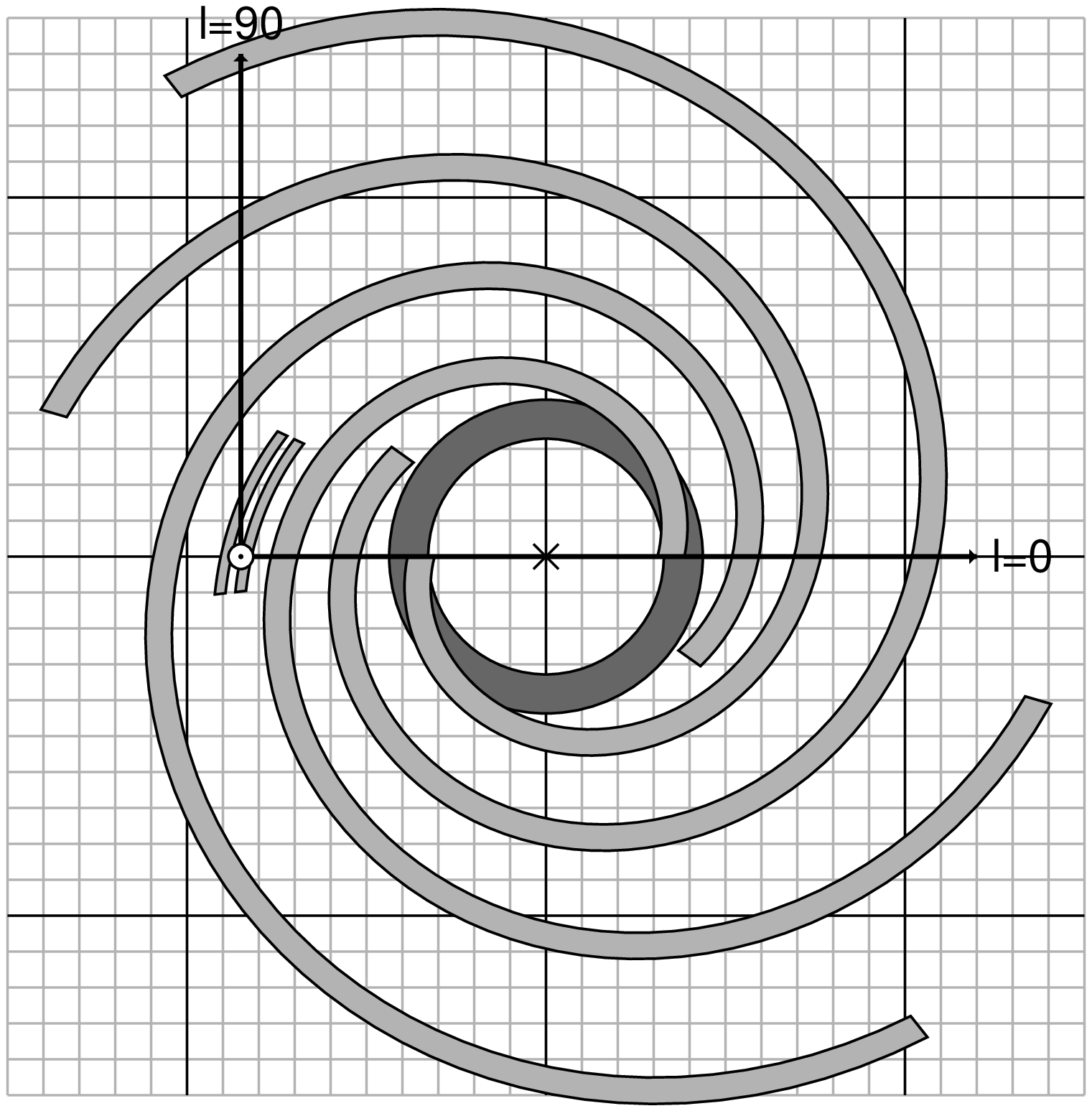}{The pattern of spiral arms and the molecular ring
in the starcount model.
Each little square is 1~kpc wide.\label{fig:spiral}}

\myfigure{elias.ps}{
Comparison of model predictions with the observed cumulative star
counts at the NGP \citep{El78}.  The total counts (solid), disk
component (dashed), spiral arm component (dot-dashed), and halo (dotted)
are all shown.  The dominance of the disk component which holds for $|b|
\geq 20^\circ$ is readily apparent.  The ring and bulge components are
absent because of the high galactic latitude considered in this plot.
\label{fig:elias}}

\myfigure{2MASS.ps}{Comparison of the differential 
star counts
from the 2MASS survey at 2.2 \um\ to the \citet{Co94} starcount model
in six different fields.
Galaxy counts become larger than the high $b$ starcounts for
$K \geq 15.5$ \protect\citep*{GCW93}.
\label{fig:2MASS}}

\myfigure{eaton.ps}{Comparison of the cumulative star counts in the
galactic plane from \E\ with the \citet{Co94} model star counts.  The
model is consistent with the data with the exception of the galactic
center.\label{fig:eaton}}

\myfigure{2obs_diff_6.ps}{Comparison of the \citet{Co94} model counts with 
high galactic latitude observations over a wide range of
magnitudes. The 2MASS points are the average of all 6 fields.
\label{fig:3obs_diff}}

\myfigure{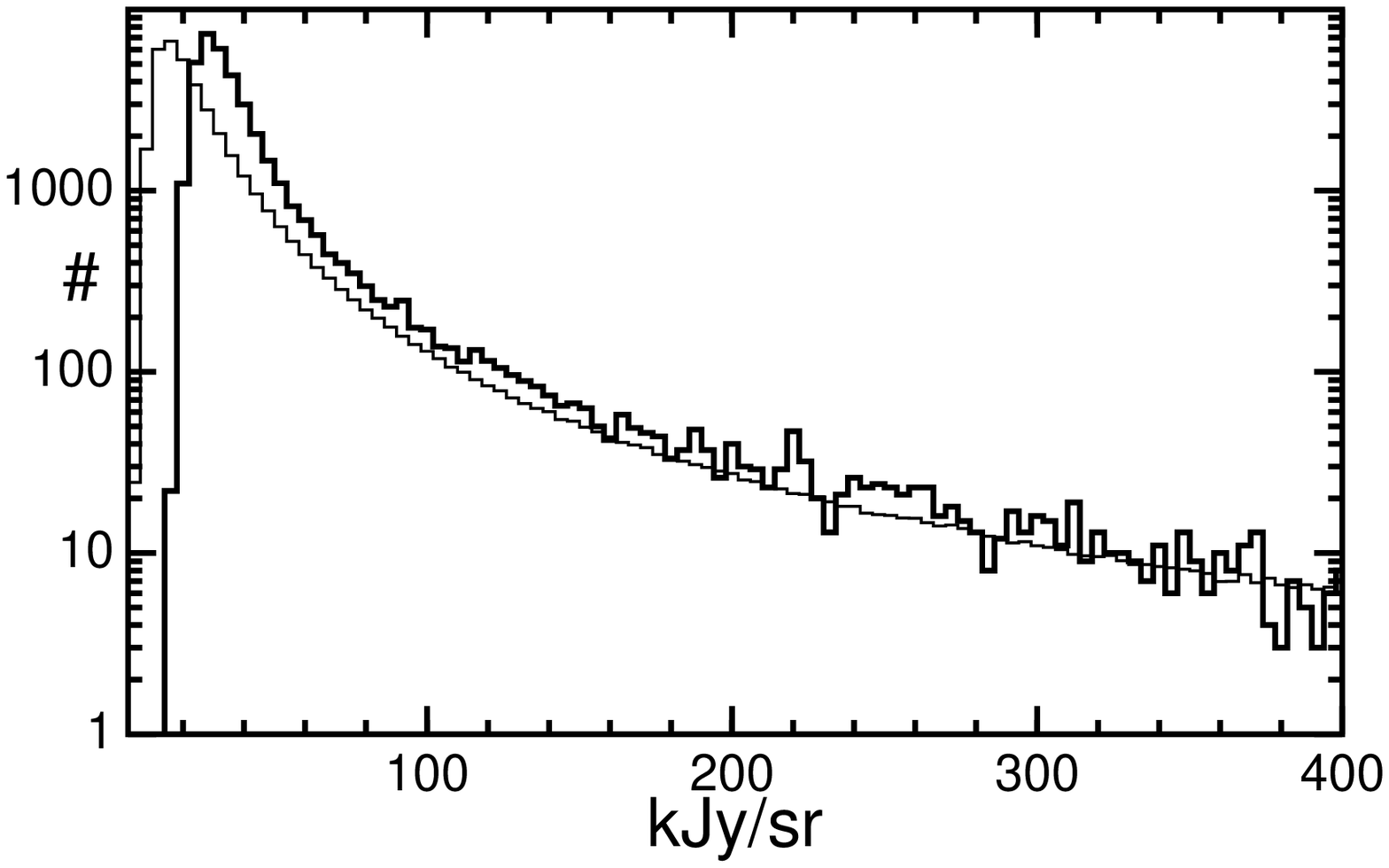}{The DIRBE and model histograms of number of pixels 
receiving a given flux.  The heavy line shows the real data for
the region with $\sin|b| > 0.9$ at 3.5 \um\ while 
the thin line shows the predicted 
histogram computed for 100 times more pixels than the real 
data and scaled down by a factor of 100.
The offset between the two histograms is the CIRB.
\label{fig:hist_eye}}

\myfigure{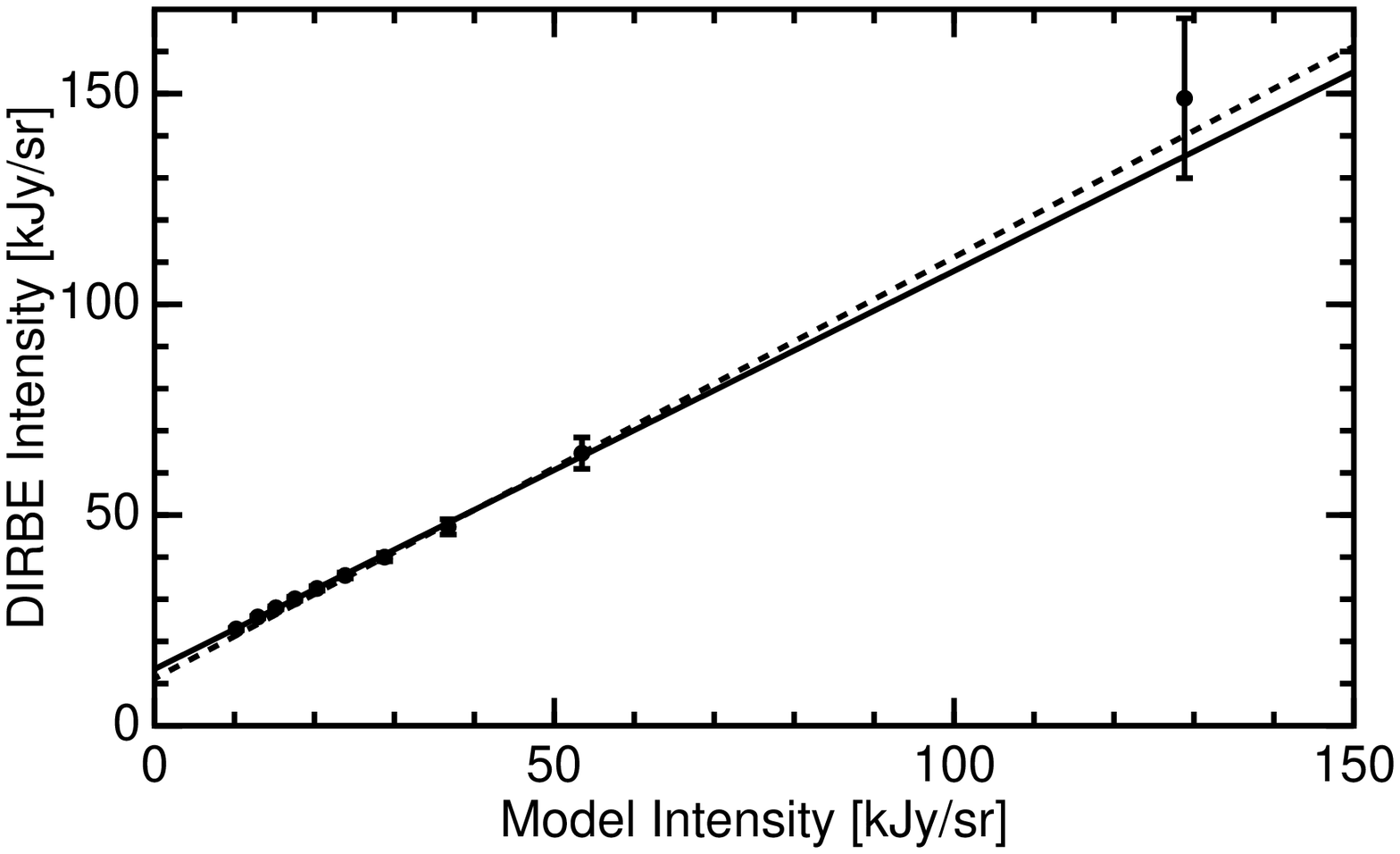}{DIRBE and model histogram values at 
the 5$^{th}$, 15$^{th}$, \ldots 95$^{th}$ \%-tiles
for the region with $\sin|b| > 0.9$ at 3.5 \um.
Two fits are shown: one with the slope fixed to unity [dashed],
and the second with the slope as a free parameter [solid].
\label{fig:fit}}

\myfigure{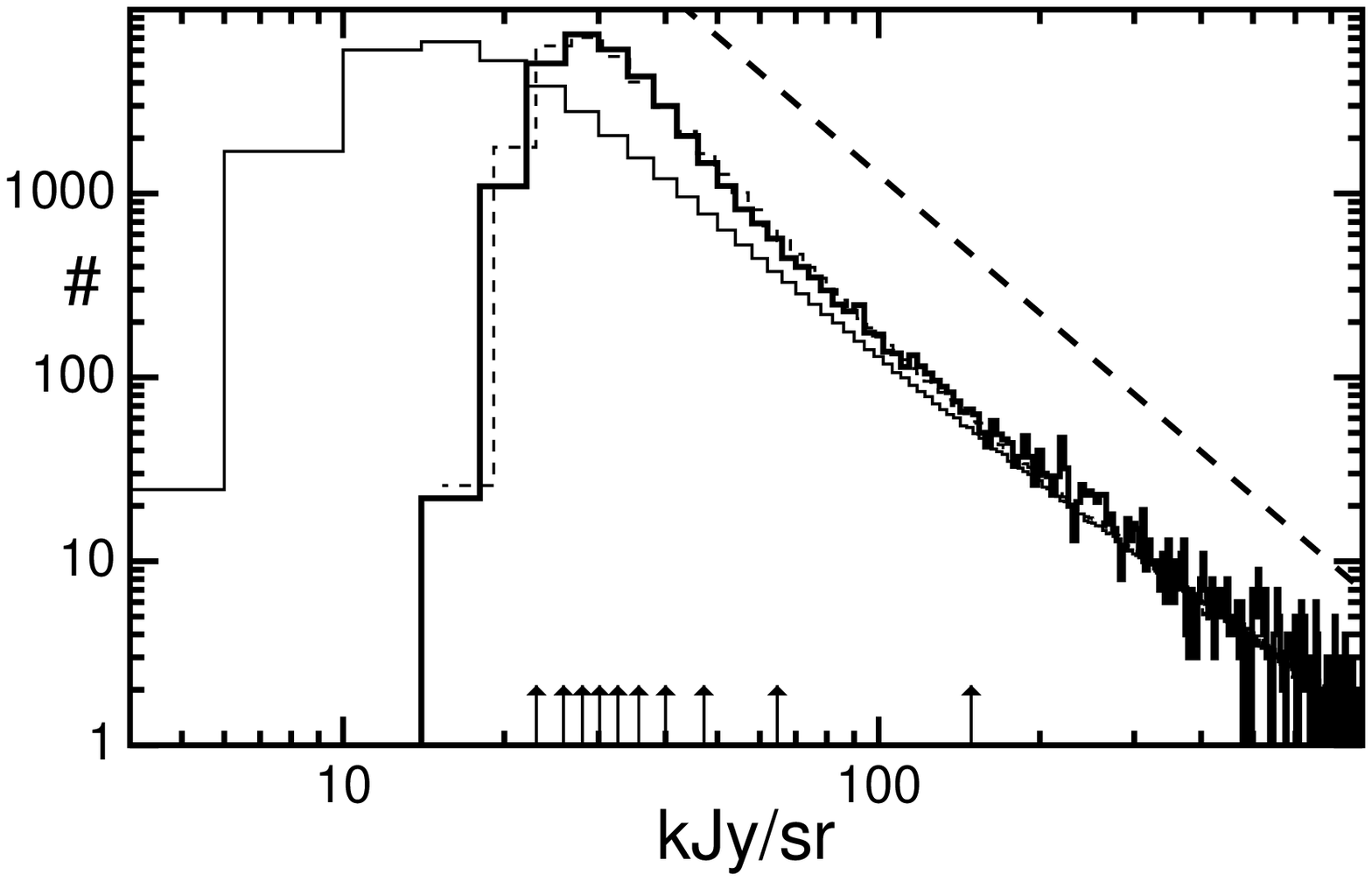}{The DIRBE and model histograms of number of pixels 
receiving a given flux in the region with $\sin|b| > 0.9$ at 3.5 \um.  
Both the original model histogram and the best fit with slope 0.9485
and intercept 13.42 \kJysr\ are shown.
The arrows along the x-axis show the positions of the
percentiles used in fitting.
The dashed straight line with a slope of -5/2 shows a Euclidean source count.
\label{fig:hist_ks}}

\myfigure{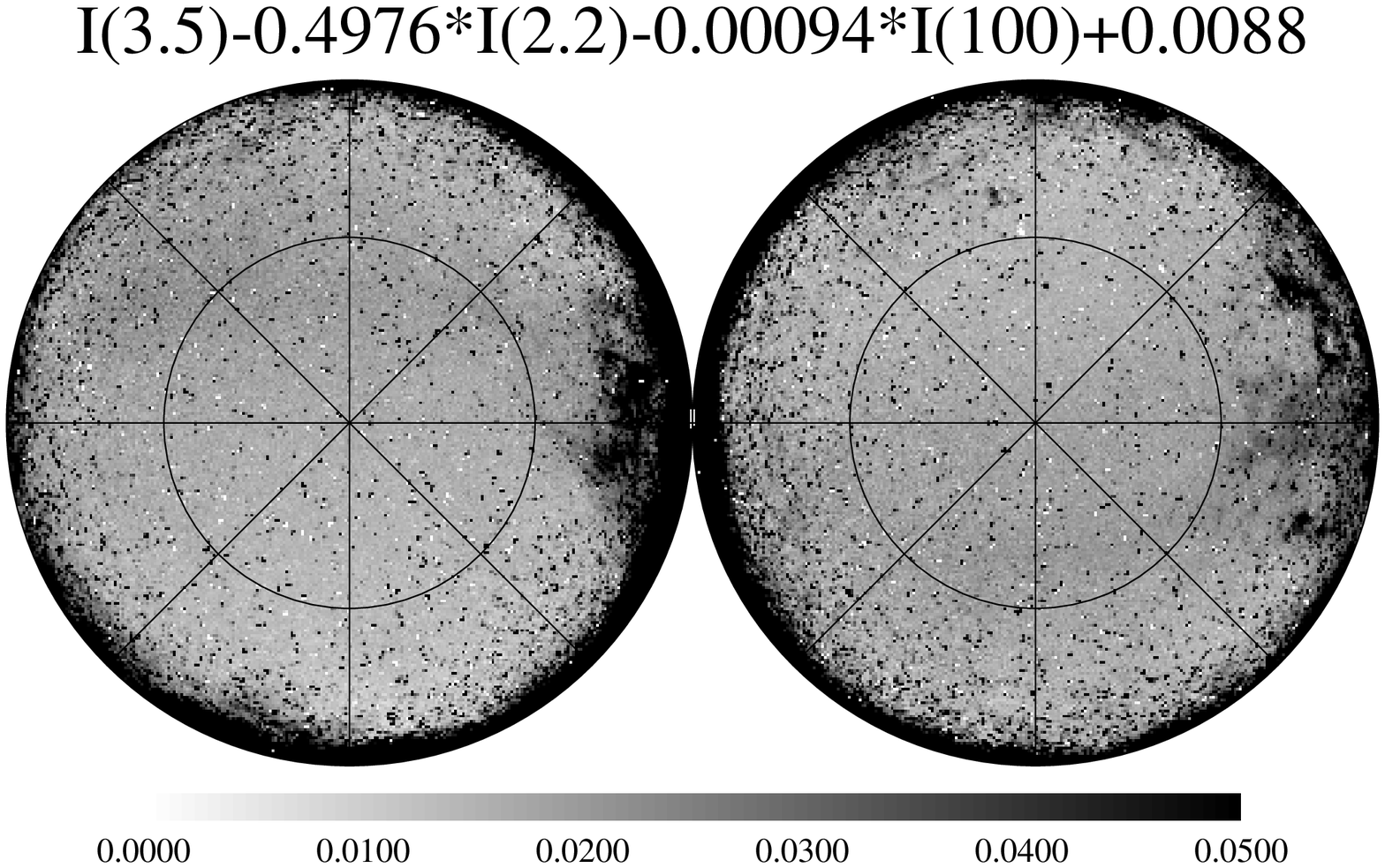}{The residual at 3.5 \um\ after 
subtracting a fraction of the 2.2 \um\ map to
remove stars and a fraction of the 100 \um\ map to remove ISM dust emission.
$l = 0^\circ$ is at the center of the figure, $b=90^\circ$ is the center of 
the left circle, and $l=90^\circ$ is toward the bottom of both circles.
Intensities are given in \MJysr.
\label{fig:L-K-100}}

\myfigure{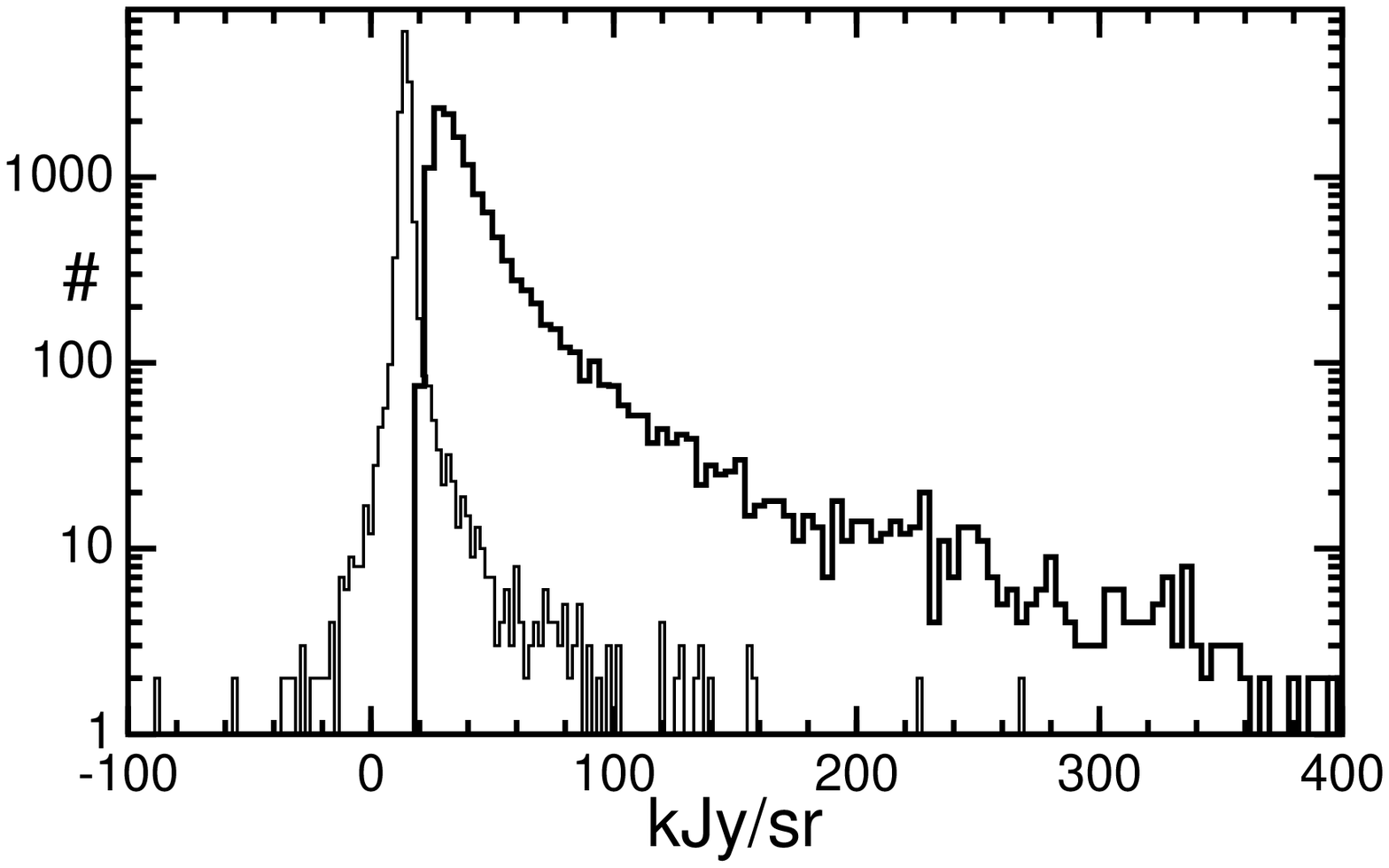}{The histogram of the zodi-subtracted 3.5 \um\ map 
in the region with
$b > 45^\circ$ and $\beta > 45^\circ$ (thick line) compared to the histogram
of the residual map from Figure~\protect\ref{fig:L-K-100} in the same region
(thin line).
\label{fig:hist-2}}

\end{document}